\documentclass[aps,10pt,prl,twocolumn,groupedaddress,nofootinbib,notitlepage,preprintnumbers]{revtex4-2}
\usepackage[utf8]{inputenc}
\usepackage{hyperref}
\usepackage{xcolor}
\usepackage{graphicx}
\usepackage{amsmath,amssymb}
\usepackage{bm}
\usepackage[shortlabels]{enumitem}
\begin{document}
\title{Were recently reported MHz events planet mass primordial black hole mergers?}

\author{\textsc{Guillem Dom\`enech$^{a}$}}
    \email{{domenech}@{pd.infn.it}}

\affiliation{$^{a}$ \small{INFN Sezione di Padova, I-35131 Padova, Italy}}

\begin{abstract}
A bulk acoustic wave cavity as high frequency gravitational wave antenna has recently detected two rare events at $5.5$MHz. Assuming that the detected events are due to gravitational waves, their characteristic strain amplitude lies at about $h_c\approx 2.5 \times 10^{-16}$. While a cosmological signal is out of the picture due to the large energy carried by the high frequency waves, the signal could be due to the merging of two planet mass primordial black holes ($\approx 4\times 10^{-4} M_\odot$) inside the Oort cloud at roughly $0.025$ pc ($5300$ AU) away. In this short note, we show that the probability of one such event to occur within this volume per year is around $1:10^{24}$, if such Saturn-like mass primordial black holes are $1\%$ of the dark matter. Thus, the detected signal is very unlikely to be due the merger of planet mass primordial black holes. Nevertheless, the stochastic background of saturn mass primordial black holes binaries might be seen by next generation gravitational wave detectors, such as DECIGO and BBO.
\end{abstract}

\maketitle

\section{Introduction \label{sec:Intro}}

Gravitational waves offer new means to probe the unexplored universe and may lead to new discoveries in cosmology and astrophysics. Gravitational wave interferometers present the opportunity to test compact objects, such as black holes, with masses ranging from tenth of solar masses at frequencies $10-1000\,{\rm Hz}$, the LIGO range, to million solar masses at frequencies $10^{-4}-10^{-2}\,{\rm Hz}$, the LISA range. Pulsar timing arrays may probe supermassive black holes with tenths of billions of solar masses at frequencies $10^{-9}-10^{-7}\,{\rm Hz}$. These are also very interesting windows for cosmology as it gives access to extraordinary physics in the early universe when the temperature was around $0.1-10^{11}\,{\rm GeV}$. Such powerful events could be (see \cite{Caprini:2018mtu} for a review) first order phase transitions, cosmic strings, etcetera. It also provides the means to test the abundance of primordial black holes \cite{Nakamura:1997sm,Khlopov:2008qy,Saito:2008jc,Saito:2009jt,Mandic:2016lcn,Wang:2016ana,Bird:2016dcv,Sasaki:2016jop,Ali-Haimoud:2017rtz,Sasaki:2018dmp,Wang:2019kaf,Kohri:2020qqd,Yuan:2021qgz}, that is black holes that formed in the early universe, with the gravitational waves due to the mergers of binaries and the secondary gravitational waves produced around the time of primordial black hole formation. See \cite{Sasaki:2018dmp} for a review on the former and \cite{Yuan:2021qgz,Domenech:2021ztg} for reviews on the latter.

In principle, cosmology as well as exotic astrophysical objects, such as bosons stars, might also produce gravitational waves with much higher, e.g. from MHz to GHz. Interestingly, as no known astrophysical source emits such high frequency waves, this is a unique window to test early universe physics \cite{Aggarwal:2020olq}. For example, high frequency cosmological gravitational waves could have been produced during preheating and phase transitions \cite{Liu:2017hua,Liu:2018rrt,Cai:2021gju}. Resonant mass detectors are a promising way to detect such high frequency gravitational waves \cite{Aggarwal:2020olq,Goryachev:2014yra}. These high frequency waves can excite the vibrational eigenmodes of, e.g., a spherical mass which are then translated and amplified to electric signals. However, the main obstacle is that due to the large energy carried by such high frequency waves, the characteristic strain of a cosmological stochastic gravitational wave background must be minuscule.\footnote{An estimation tells us that \cite{Maggiore:1900zz}
\begin{align*}
\Omega_{\rm GW,0}h^2&=\frac{4\pi^2}{3H_{100}^2}f_{\rm GW}^2h_c^2\\&\approx1.3\times10^{-6}\left(\frac{f_{\rm GW}}{\rm MHz}\right)^2\left(\frac{h_c}{10^{-27}}\right)^2\nonumber\,.
\end{align*}
Compare this value with the current bounds from Big Bang Nucleosyntesis (BBN) which are $\Omega_{\rm GW,BBN,0}h^2<1.8\times 10^{-6}$ \cite{Aghanim:2018eyx,Caprini:2018mtu}. This means that in order to be competitive, we need a very high sensitivity of the resonant mass detectors to reach $h_c\lesssim 10^{-27}$.} Nevertheless, we might hope to detect nearby astrophysical sources.

Recently, there has been a detection of two rare events at around $5\,{\rm MHz}$ from a bulk acoustic wave antenna \cite{Goryachev:2021zzn}, which might be due to gravitational waves, though further confirmation is needed. Note that the MHz interferometer at Fermi Lab, called Holometer \cite{Holometer:2016qoh}, did not detect any signal in the range of $1-13\,{\rm MHz}$ in their first run. However, it was not sensitive to fast transient signals \cite{Goryachev:2021zzn}. In any case, assuming that the events of \cite{Goryachev:2021zzn} are due to gravitational waves, the corresponding characteristic strain is around $h_c\sim 2.5\times 10^{-16}$. As already noted in \cite{Goryachev:2021zzn} the signal could be due to the merger of small compact objects such as primordial black holes with masses of about $10^{-4}M_\odot$, where $M_\odot\approx 2\times 10^{33}\,{\rm g}$ is a solar mass, and at a distance of roughly $0.01\,{\rm pc}$. In this note, we investigate this claim in more detail to show that the probability of detecting the gravitational waves from the merger of a primordial black hole binary with such small masses is extremely small. Nevertheless, we may hope to detect the stochastic signal due to the merger of sub-solar and planet-mass primordial black hole binaries in the far future \cite{Wang:2019kaf,Mukherjee:2021ags,Mukherjee:2021itf}. Throughout the paper we work in units where $c=\hbar=1$. Also, whenever needed we use the cosmological parameters provided by the Planck collaboration \cite{Aghanim:2018eyx}.

\section{Gravitational waves from primordial black holes mergers\label{sec:1}}

Binaries emit gravitational waves as they inspiral with a final burst when the two black holes merge \cite{Maggiore:1900zz}. The largest gravitational wave amplitude is generated in the latest stages of the merger at the so-called chirp. We can estimate the chirp frequency by using the frequency associated to the radius of the last Innermost Stable Circular Orbit (ISCO). In this way we have that \cite{Maggiore:1900zz}
\begin{align}\label{eq:fmaxisco}
f_{\rm GW,max}&\approx 2f_{\rm ISCO}\nonumber\\&\approx4.4\,{\rm MHz}\left(\frac{10^{-3}M_\odot}{m_1+m_2}\right)\,,
\end{align}
where $m_1$ and $m_2$ are the individual masses of the black holes. From now on, and otherwise stated, we assume for simplicity that the primordial black hole mass spectrum is monochromatic and so $m_1=m_2=M_{\rm PBH}$ where PBH stands for primordial black hole. The conclusion does not change significantly if $m_1$ and $m_2$ differ slightly. From \eqref{eq:fmaxisco} we see that in order to explain the detected $5.5\,{\rm MHz}$ events we need $M_{\rm PBH}\sim 4\times 10^{-4} M_\odot$. These primordial black holes have a Schwarzschild radius\footnote{In more detail the Schwarzschild radius $r_s$ is given by
\begin{align}
r_s=2GM_{\rm PBH}\approx 300 \,{\rm cm} \left(\frac{M_{\rm PBH}}{10^{-3}M_\odot}\right)\,.
\end{align}} 
of $120 \,{\rm cm}$. So, they are the size of a giant Pilates ball.

However, the amplitude of the gravitational waves generated by such very small objects is rather tiny and, as we shall see, they must have merged close to the edge of the solar system, inside the Oort cloud. Since they merged in the local neighbourhood, we shall neglect the cosmological expansion and use that the characteristic strain of a inspiraling binary is given by \cite{Maggiore:1900zz}
\begin{align}
h_c&\approx \frac{4}{r}\left(GM_c\right)^{5/3}\left(\pi f_{\rm GW}\right)^{2/3}\,.
\end{align}
Using the chirp frequency estimate in \eqref{eq:fmaxisco} we find that
\begin{align}
h_{c,\rm max}&\approx \frac{4}{r}\left(GM_c\right)^{5/3}\left(\pi f_{\rm GW,max}\right)^{2/3}\nonumber\\&\approx
2\times 10^{-17}\left(\frac{M_c}{10^{-3}M_\odot}\right)\left(\frac{r}{\rm pc}\right)^{-1}\,.
\end{align}
Thus, in order to match the amplitude of the possibly detected signal of about $h_{c,\rm max}\approx 2.5\times 10^{-16}$ \cite{Goryachev:2021zzn} the primordial black holes merged at $r\sim 2.6\times 10^{-2}\,{\rm pc}$. Note that taking the peak strain sensitivity of $5\times 10^{-19}$ at $5{\rm MHz}$, the observable volume is around $(0.1\,{\rm pc})^3$. Now, to claim that the merger of primordial black holes is a plausible explanation we must compute the merger rate of such binaries.

\subsection{Merger rate}

To compute the rate at which primordial black holes merge within the observable volume of $(0.1\,{\rm pc})^3$, we follow references \cite{Nakamura:1997sm,Sasaki:2016jop,Ali-Haimoud:2017rtz,Sasaki:2018dmp}. In particular, using the results of \cite{Ali-Haimoud:2017rtz} we have that the current merger rate is given by
\begin{align}\label{eq:RR}
{\cal R}&\equiv \frac{dN_{\rm merge}}{dtdV}\approx  {1.5\times 10^{-18}}\text{pc}^{-3} \text{yr}^{-1}\nonumber\\&\times \frac{f_{\rm PBH}^2}{\left(f_{\rm PBH}^2+\sigma_{\rm eq}^2\right)^{21/74}}  \left(\frac{M_{\rm PBH}}{10^{-3}M_\odot}\right)^{-32/37}\,,
\end{align}
where $f_{\rm PBH}$ is the energy density fraction of primordial black holes with respect to dark matter and $\sigma_{\rm eq}^2\approx 2.5\times 10^{-5}$ is the variance of density perturbations of the dark matter not in the form of primordial black holes at matter-radiation equality. This estimate assumes that primordial black holes form randomly from the collapse of large primordial fluctuations and are uniformly distributed in the early universe  (see \cite{Sasaki:2018dmp} for a review). Note that in the recent years there have been several improvements of the PBH binary merger rate with respect to equation \eqref{eq:RR}, which take into account the torque from all PBHs, later interaction with other PBHs and accretion of surrounding matter \cite{Raidal:2018bbj,Liu:2018ess,Liu:2019rnx,Vaskonen:2019jpv,Hutsi:2020sol}. These effects tend to change the estimate \eqref{eq:RR} by an $O(1)\sim O(10)$ factor, generally suppressing the merger rate. However, since we find that the merger rate is extremely low to be able to explain the MHz events, by several orders of magnitude, we prefer to use \eqref{eq:RR} for simplicity. Any additional factors do not change the main result of this work.

From now on we will only be interested in the case where $f_{\rm PBH}>\sigma_{\rm eq}$ and so we can safely drop the $\sigma_{\rm eq}$ dependence in \eqref{eq:RR}. Assuming that $f_{\rm PBH}\sim 0.01$ so that observational constraints\footnote{Note that OGLE reported the detection of few microlensing events of earth mass objects \cite{2017Natur.548..183M} which could potentially be primordial black holes. These objects are too light to explain the MHz events.} from microlensing are satisfied \cite{Niikura:2019kqi,Carr:2020gox,Carr:2020xqk,Green:2020jor}, we find that ${\cal R}\lesssim 5\times 10^{-24}$ for $M_{\rm PBH}=4\times 10^{-4} M_\odot$ within $(0.1\,{\rm pc})^3$. Thus, we see that the probability that such rare events are due to the merger of Saturn-like mass primordial black holes is extremely low. Nevertheless, even if we have been very lucky to detect two of such mergers, we also must take into account the stochastic background of gravitational waves generated by the large number of mergers scattered across the universe.

\subsection{Stochastic gravitational wave background}

In the scenario under study there would be many primordial black hole binaries merging since their formation throughout the universe's history. The spectral density of the cumulus of inspiral and mergers is calculated according to \cite{LIGOScientific:2016fpe,Mandic:2016lcn,Wang:2016ana,Wang:2019kaf,Kohri:2020qqd}
\begin{align}\label{eq:omegastochastic}
\Omega_{\rm GW}=\frac{1}{3H_0^2M_{\rm pl^2}}\int_0^{\frac{f_{\rm cut}}{f}-1}dz\frac{{\cal R}(z)}{(1+z)H(z)}f\frac{dE_{\rm GW}}{df_s}\,,
\end{align}
where $f$ is the measured gravitational wave frequency, $z$ is the redshift, $H(z)$ is the Hubble parameter or expansion rate, $H_0=H(0)$ is the Hubble parameter today, $M_{\rm pl}^2=(8\pi G)^{-1}$, ${\cal R}(z)$ is the merger rate of primordial black holes in terms of $z$ which can be found in \cite{Ali-Haimoud:2017rtz}, and ${dE_{\rm GW}}/{df_s}$ is the energy emitted per binary per frequency where $f_s=(1+z)f$ is the frequency at the source. The spectrum is cut-off at $f_{\rm cut}$ where the binary completely merged and no further gravitational waves are emitted. Using the formulas in the appendix of \cite{Wang:2019kaf} we find that
\begin{align}
f_{\rm cut}\approx 11\, {\rm MHz} \left(\frac{M_{\rm PBH}}{10^{-3}M_\odot}\right)^{-1}\,.
\end{align}
We also have from \cite{Wang:2019kaf} that
\begin{align}
&\frac{dE_{\rm GW}}{df_s}\approx\frac{\pi^{2/3}}{3G}\left(GM_c\right)^{5/3}\nonumber\\&\times\left\{
\begin{aligned}
&f_s^{-1/3} & f_s<f_1\\
&\omega_1 f_s^{2/3} & f_1<f_s<f_2\\
&\omega_2 \frac{\sigma^4f_s^{2}}{\left(\sigma^2+(f_s-f_2)^2\right)^2} & f_2<f_s<f_{\rm cut}\\
& 0 & f_{\rm cut}<f_s
\end{aligned}
\right.\,,
\end{align}
where $\omega_1$ and $\omega_2$ are found by continuity and \cite{Ajith:2009bn,Wang:2019kaf}
\begin{align}
f_1&\approx 3.6\, {\rm MHz} \left(\frac{M_{\rm PBH}}{10^{-3}M_\odot}\right)^{-1}\,,\\ f_2&\approx 8.2\, {\rm MHz} \left(\frac{M_{\rm PBH}}{10^{-3}M_\odot}\right)^{-1}\,,\\
\sigma&\approx 1.9\, {\rm MHz} \left(\frac{M_{\rm PBH}}{10^{-3}M_\odot}\right)^{-1}\,.
\end{align}

We can have a rough estimation of the magnitude of the stochastic background due to the binaries as follows. We note that most of the energy is carried away by the high frequency waves emitted by the final gravitational wave bursts of the primordial black hole binary mergers. Most of the high frequency gravitational waves must have been emitted in the nearby universe, so that they are not significantly affected by the cosmological expansion. This is confirmed by a numerical calculation \cite{Wang:2019kaf} where it is found that the peak of the gravitational wave spectrum is close to the binary cut-off frequency, which is around $5$ times higher than the frequency corresponding to the ISCO \eqref{eq:fmaxisco}. As a rough approximation we can thus evaluate \eqref{eq:omegastochastic} at low redshift and at $f_s\sim f=\alpha f_{\rm cut}$ with $z=\alpha^{-1}-1<1$, which yields
\begin{align}\label{eq:omegastochastic2}
\Omega^{\rm peak}_{\rm GW}\sim \frac{\left(1-\alpha\right)}{3H_0^2M_{\rm pl^2}}\frac{{\cal R}}{H_0}\frac{dE_{\rm GW}}{df}\bigg|_{\alpha f_{\rm cut}}\,.
\end{align}
The maximum of such function in terms of $\alpha$ is found at $\alpha\sim 0.73$, which is close enough to $f_2$ but still $\alpha f_{\rm cut}>f_2$. This implies that $z\sim 0.37$ and so our approximation should give a fair enough estimate. With these values we find that
\begin{align}\label{eq:omegastochastic3}
\Omega^{\rm peak}_{\rm GW}\approx 6.1\times 10^{-9}\left(\frac{M_{\rm PBH}}{10^{-3}M_\odot}\right)^{5/37}\left(\frac{f_{\rm PBH}}{0.01}\right)^{53/37}\,,
\end{align}
which is close enough to the numerical results shown in \cite{Wang:2019kaf}.\footnote{Note that the magnitude of the gravitational wave peak power \eqref{eq:omegastochastic3} is similar to the total power emitted by the inspiral phase today, i.e. $\rho_{\rm GW}^{\rm ins}\sim {\cal R}\Delta E_{\rm max}/H_0$ where $\Delta E_{\rm max}={\pi^{2/3}}\left(GM_c\right)^{5/3}f_{\rm GW, max}^{2/3}/(3G)$ \cite{Maggiore:1900zz}.}
Then, we know that the spectrum roughly decays as $f^{2/3}$ for $f<\alpha f_{\rm cut}$ and is practically zero for $f>\alpha f_{\rm cut}$ \cite{Wang:2019kaf}. Interestingly, while this type of gravitational wave background \eqref{eq:omegastochastic3} may be far from future BBN bounds, e.g. by CMB-S4 \cite{Abazajian:2016yjj}, and resonant mass detectors, the low frequency tail of the spectrum \eqref{eq:omegastochastic} for $M_{\rm PBH}\sim 10^{-4}-10^{-3}M_\odot$ and $f_{\rm PBH}\sim 0.01$ enters the observable range of future gravitational wave interferometers such as DECIGO \cite{Seto:2001qf,Yagi:2011wg,Kawamura:2020pcg}, LISA \cite{Audley:2017drz},  BBO \cite{Moore:2014lga}, Einstein Telescope \cite{Maggiore:2019uih}, AEDGE \cite{AEDGE:2019nxb} and Cosmic Explorer \cite{LIGOScientific:2016wof}, as was shown in \cite{Wang:2019kaf,Pujolas:2021yaw}.\footnote{For future prospects on the detectability of the stochastic background due to the merger solar and sub-solar mass PBH binaries, $M_{\rm PBH}\sim O(0.1)-O(10)M_\odot$, see \cite{Mukherjee:2021ags,Mukherjee:2021itf}.} For example, let us extrapolate \eqref{eq:omegastochastic4} to a frequency of $f\sim 0.1\,{\rm Hz}$. This gives
\begin{align}\label{eq:omegastochastic4}
&\Omega_{\rm GW}(0.1{\rm Hz})\sim \Omega_{\rm GW}^{\rm peak}\left(\frac{f}{\alpha f_{\rm cut}}\right)^{2/3}\bigg|_{f=0.1\,{\rm Hz}}\nonumber\\& \approx 3\times 10^{-14}\left(\frac{M_{\rm PBH}}{10^{-3}M_\odot}\right)^{89/111}\left(\frac{f_{\rm PBH}}{0.01}\right)^{53/37}\,.
\end{align}
Let us stress that these are order or magnitude estimates and that they differ from the numerical calculation by a factor $O(1)$. For instance, \eqref{eq:omegastochastic4} is found to be roughly a factor $2$ smaller than the results of \cite{Wang:2019kaf}. The DECIGO and BBO peak sensitivities at $0.1\,{\rm Hz}$ are expected to be around $\Omega^{\rm DECIGO}_{\rm GW}\sim 10^{-14}$ and $\Omega^{\rm BBO}_{\rm GW}\sim 10^{-15}$. Thus, any MHz gravitational wave signal due to Saturn-like mass primordial black holes which make up for $O(1\%)$ of dark matter must have an stochastic background signal in principle detectable by future gravitational wave detectors.

\section{Conclusions} 

It is an exciting time for cosmology and astrophysics as new gravitational wave data is becoming available. Recently, two rare events at frequencies of $5.5\,{\rm MHz}$ have been detected by a bulk acoustic detector \cite{Goryachev:2021zzn}. If these events are shown to be gravitational waves, they would be an important hint of exotic physics, either from cosmological or astrophysical origin. In this note we have shown that, if indeed it is the case, the $5.5\,{\rm MHz}$ gravitational waves with characteristic strain $h_c\sim 2.5\times 10^{-16}$ are very unlikely to be due to the merger of Saturn-like mass primordial black hole binary, with $M_{\rm PBH}\sim 4\times 10^{-4} M_\odot$. The probability of detecting a single such event is less than $1:10^{24}$. This renders this scenario practically implausible unless these primordial black holes are extremely clustered around us. Even so, microlensing observations would probably rule out this option. Thus, Saturn-like mass primordial black holes cannot account for the rare events. Nevertheless, the stocastic background of saturn mass primordial black hole binaries is in principle detectable by DECIGO and BBO if primordial black holes make up to $O(1\%)$ of dark matter.
\newpage

\textbf{Note:} After submission, another work by Lasky and Thrane \cite{Lasky:2021naa} appeared to cast doubts on the detection of the MHz gravitational waves \cite{Goryachev:2021zzn}. They argue that a large gravitational wave memory signal should have been seen by LIGO. The non-detection of such memory signals apparently rule out the gravitational wave explanation for the MHz events of \cite{Goryachev:2021zzn}.

\section*{Acknowledgments} 
We would like to thank Misao Sasaki for useful discussions and feedback. We would also like to thank Paul Lasky and Eric Thrane for sharing an earlier version of their paper. G.D. as a Fellini fellow was supported by the European Union’s Horizon 2020 research and innovation programme under the Marie Sk{\l}odowska-Curie grant agreement No 754496.

\bibliography{bibliographyreview.bib}

\end{document}